\documentclass[reprint,amsmath,amssymb,aps,]{revtex4-1}
\usepackage{graphicx}
\usepackage{dcolumn}
\usepackage{bm}

\newcommand{\nn}{\nonumber}
\newcommand{\beq}{\begin{eqnarray}}
\newcommand{\enq}{\end{eqnarray}}
\newcommand{\bear}{\begin{array}}
\newcommand{\edar}{\end{array}}
\newcommand{\beqar}{\begin{eqnarray}}
\newcommand{\enqar}{\end{eqnarray}}

\newcommand{\bvp}{\bm{v_p}}
\newcommand{\bv}{\bm{v}}

\newcommand{\bom}{\bm{\omega}}
\newcommand{\ddt}{\frac{d}{dt}}
\newcommand{\pdt}{\frac{\partial}{\partial t}}

\newcommand{\nr}{\nn \\}

\begin{document}
\preprint{}

\title{Spin reversal of a rattleback with viscous friction}

\author{Hiroshi Takano}
\affiliation{%
 Joetsu university of education\\
 943-8512 1 Yamayashiki Joetsu Niigata Japan
}%
\email{takano@juen.ac.jp}

\date{December 21, 2011}

\begin{abstract}
Effective equation of motion of a rattleback is obtained from the basic equation of motion
 with viscous friction depending on  slip velocity.
This effective equation of motion is used to estimate the number of spin reversals
and the rattleback shape that causes the maximum number of spin reversals.
These estimates are compared with numerical simulations based on the basic equation of motion.
\end{abstract}

\maketitle
%
\section{Introduction}
\begin{figure}[h]
  \begin{center}
    \includegraphics[width=0.45\textwidth]{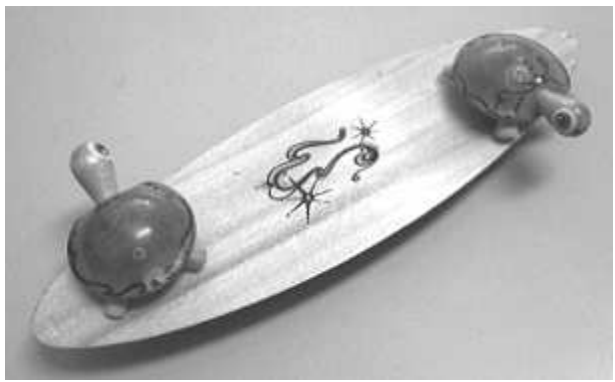}
    \caption{A Russian rattleback toy called  stubborn tortoise.
    \label{kame}
    }
  \end{center}
\end{figure}

%
A rattleback, also known as  a celt or wobblestone, is a type of mechanical top
 with the curious property of spin asymmetry.
Nowadays, there are different varieties of rattlebacks. 
Fig. \ref{kame} shows a Russian rattleback toy called  stubborn tortoise.
When the rattleback is spun in the clockwise direction which the tortoise has turned to,
 it continues to spin clockwise
until it slows to a stop. However, when the rattleback is spun in the anticlockwise
direction, a self induced oscillation occurs, and
the spin slows down and eventually its direction is reversed.
There is an inertial asymmetry, because the tortoise's center of mass is shifted 
from the principal axes of the body surface ellipsoid.
This inertial asymmetry is responsible for spin reversal.

Many analyses and simulations have attempted to 
explain the dynamics of rattlebacks during the last century.

The first scientific paper on rattlebacks appeared in 1898 by Walker\cite{gtwalker}.
Assuming dissipation-free rolling without slipping, he obtained  linearized
equations of motion for contact point coordinate variables and spin variables.
He analyzed the instability of the spin magnitude and direction by studying a
characteristic equation and showed the relationship between the direction of spin and the oscillation.
For a realistic case, the spin $n$ is small; therefore,
 his analysis can explain the behavior of a rattleback.

In 1983, Pascal\cite{pas}, using the same assumptions as Walker\cite{gtwalker}, derived effective 
equations of motion for the slowly varying mode by using the method of averaging
and clarified a rattleback's reversal mechanism.

During the same period, Markeev\cite{mar} obtained results similar to that of Pascal\cite{pas}.
He derived two conservation laws and provided a comprehensible explanation of a rattleback's reversal.
Their results were extended to second-order averaging 
by Blackowiak, Rand and Kaplan\cite{black}.

Moffatt and Tokieda\cite{mt} presented a physically transparent derivation of
the effective equations of motion similar to  Markeev's derivation.

In 1986, Bondi\cite{bondi} extended Walker's\cite{gtwalker} results to understand how  spin evolves 
for a wide range of geometric and inertial parameters of the body.

For a no-slip dissipation-free case, numerical simulations \cite{kl}\cite{ll}\cite{gh}
 showed that infinite spin reversals occur; however, a real rattleback has finite spin reversals 
 because of energy loss by slip friction. 
 Thus, it is important to analyze the dynamics of rattlebacks with slip friction.

Furthermore, Karapetian\cite{kara} discussed the stability of rotation 
of a heavy asymmetric rigid body on a horizontal plane(celtic stone); however, he did not
discuss the number of spin reversals.
Garcia and  Hubbard\cite{gh} discussed the limitations of a no-slip case and 
analyzed the effects of dissipation. 
They derived augmented equations of motion incorporateing lumped models for aerodynamic effects,
spinning torque and slipping torque due to Coulomb friction force of slip velocity.
In reality, the  contact with the horizontal plane is not a point but an area,
relative angular motion between the surfaces cause these spinning torque and slipping torque.
These equations were solved numerically.
Because these equations were too complicated, they presented a simplified model of spin.
The spin model derived by considering  energy in the spin, oscillation, 
and dissipation was successful in explaining spin dynamics; however, 
this model was not derived from equations of motion.
Thus, the effect of the slip velocity is not clear. Furthermore, they did not discuss the number
of spin reversals and its relationship to the rattleback shape.

The observation of actual behavior of rattlebacks leads to three questions.

First, for the no-slip case, if the initial spin value $n_0$ is small, 
the rattle oscillation becomes large and  spin reversal occurs in theory. 
In contrast, for the slip case,
if $n_0$ is small, the rattle oscillation does not increase, and  spin reversal does not occur.
In this case, a critical value of the initial spin $n_c$ seems to exist,
 above which the spin reversal occurs.
What is the value of $n_c$?

Second, a real rattleback has a finite number of spin reversals because of energy loss by 
slip friction. When the value of the coefficient of friction is known, how many times
does the rattleback reverse and how does the number of spin reversals depend on 
the coefficient of friction?

Third, it seems that the number of spin reversals depends on 
the rattleback shape.
One rattleback may reverse only once, whereas another may  reverse
as many as three times. Given that the lower surface of rattleback is defined by an ellipsoid
 $\frac{x^2}{a^2}+\frac{y^2}{b^2}+\frac{z^2}{c^2}=1$, where $a>b>c=1$,
if $a\gg b$, the oscillations in the $x$ direction become large and spin reversal occurs rapidly.
After one reversal, the oscillations start in the $y$ direction but do not become as large, 
and spin reversal occurs slowly.
In this situation, the number of spin reversals is small.
In contrast, if $a=b$, the rattleback is a disk, and spin reversals does not occur.
Thus, it seems that  a critical ratio of $a$ and $b$ give the maximum number of spin reversals.
What is this value?

In this paper, to answer the above questions, I consider the basic equations of motion
 containing sliding friction and obtain linearized equations of motion containing slip velocity. 
I derive the effective equations of motion by approximating this velocity.
By applying the same averaging method as that used by  Pascal\cite{pas}, differential
equations of slowly changing variables are obtained.
These differential equations are used to analyze the relationship among the number of spin reversals,
the coefficient of friction, and the rattleback shape.
The spin behavior obtained from the simulation of the effective equations of motion
is compared with that obtained from the simulation of basic equations of motion.

%
%
\section{Basic equations of motion for a rattleback with  viscous friction}\label{basiceq}
\begin{figure}[h]
  \begin{center}
    \includegraphics[width=0.45\textwidth]{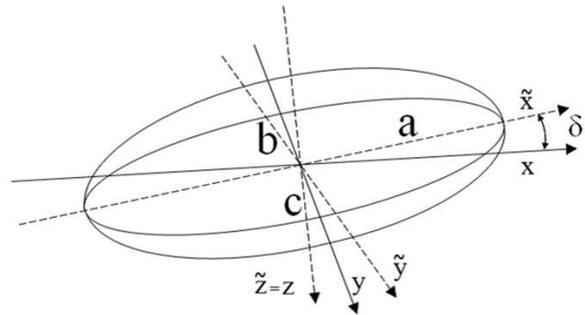}
    \caption{Body axes  $\tilde{x}$, $\tilde{y}$, and $\tilde{z}$ with lengths
     $a$, $b$, and $c$, respectively, principal inertia axes $x$, $y$, $z$ and  angle $\delta$.
    \label{xyzXYZ}
    }
  \end{center}
\end{figure}

%
%
In this paper, a rattleback is considered to be a uniform ellipsoid of mass $m$
with a  smooth lower surface such that $a>b>c$ as follows:
\beq
\frac{\tilde{x}^2}{a^2}+\frac{\tilde{y}^2}{b^2}+\frac{\tilde{z}^2}{c^2}&=& 1,\nn
\label{surface}
\enq
where $\tilde{x},\tilde{y}$, and $\tilde{z}$ are the body axes of the rattleback
 as shown in Fig. \ref{xyzXYZ}.
The distances of the three axes from the center of mass are $a$, $b$, and $c$, respectively.
The principal inertia axes are $x$, $y$, and $z$. The $z$-axis is  downward
and coincides with $\tilde{z}$-axis. The axes $x$- and $y$-axes are rotated by  angle $\delta$ such that
\begin{eqnarray}
\left(
\begin{array}{c}
\tilde{x}\\
\tilde{y}\\
\end{array}
\right)
&=&
\left(
\begin{array}{cc}
\cos \delta & -\sin \delta\\
\sin \delta & \cos \delta\\
\end{array}
\right)
\left(
\begin{array}{c}
x\\
y\\
\end{array}
\right),\nr
\tilde{z}&=&z. \nn
\label{XYZxyz}
\end{eqnarray}
In this paper, it is assumed that $\delta$ is small, such as $O(10^{-2})$.
The vector at the contact point $P$, $\bm{x}_p$, has components $x$, $y$, and $z$.
$\bm{u}$ is the upward unit vector at point $P$.

When the rattleback rotates, point $P$ is near the point $(0,0,c)$.
When the rattleback is oscillates, the values of $x$, $y$ are $|x|<a$ and $|y|<b$,
respectively, and $z$ can be expanded by a second-order approximation  of
$\frac{x}{a}<1$ and $\frac{y}{b}<1$ as follows:
\beq
z &\simeq & c\left( 1-\left\{ \frac{p}{2}\left(\frac{x}{c}\right)^2+q \frac{x y}{c^2}
+\frac{s}{2}\left(\frac{y}{c}\right)^2\right\} \right),\nn
\enq
where the parameters $p$, $q$, and $s$ are given by
\beq
p &\equiv& c^2 \left( \frac{\cos^2 \delta}{a^2}+\frac{\sin^2 \delta}{b^2} \right),\label{p}\\
q &\equiv& c^2 \cos \delta  \sin \delta \left( \frac{1}{b^2}-\frac{1}{a^2} \right),\label{q}\\
s &\equiv& c^2 \left( \frac{\sin^2 \delta}{a^2}+\frac{\cos^2 \delta}{b^2} \right).\label{s}
\enq

At the leading order, the following equations are obtained:
\beq
\bm{x}_p &\sim& (x,y,1), \label{xp}\\
\bm{u}  &\sim& (-px-qy,-qx-sy,-1). \label{u}
\enq
Assuming that $\delta$ is small, the principal moments of inertia
$I_{10}$, $I_{20}$, and $I_{30}$ are approximated as follows:
\beq
I_{10} \simeq m I_1,\quad
I_{20} \simeq m I_2,\quad
I_{30} \simeq m I_3,\nn
\enq
where
\beq
I_1 \equiv \frac{b^2+c^2}{5},\quad
I_2 \equiv \frac{a^2+c^2}{5},\quad
I_3 \equiv \frac{a^2+b^2}{5}.\nn
\enq

In the following equations, 
I consider the basic equations of motion for the angular momentum $\bm{L}$
and the velocity of the center of mass $\bm{v}_g$. In section \ref{sim},
numerical simulations of these basic equations of motion are performed.
The evolution of angular momentum  is governed by Euler's equation:
\beq
\frac{d}{dt}\bm{L} = \bm{x}_p \times \left( R \bm{f}+ R \bm{u}\right), \label{Ldot}
\enq
where $R\bm{u}$ is the normal reaction at $P$, and $\bm{f}$ is the slip friction force
at $P$.
The dynamics of the center of mass is governed by Newton's equation
for the center of mass velocity $\bm{v}_g$:
\beq
m\frac{d}{dt}\bm{v}_g=(R-mg)\bm{u}+R\bm{f} \label{vgdot}. \label{vgdot}
\enq
The slip velocity, $\bvp$, is related to velocities 
$\bm{v}_0 \equiv \bm{x}_p\times \omega$ and $\bm{v}_g$ as follows:
\beq
\bvp=\bm{v}_g-\bm{v}_0.\label{vgvp}
\enq
Because  $\bm{f}$, $\bvp$, and $\ddt{\bvp}$ only have components in the horizontal direction,
using Eqs. (\ref{vgdot}) and (\ref{vgvp}), the normal reaction $R$ is given by
\beq
R = m g + m \left(\ddt \bm{v}_0\right)\cdot \bm{u} \label{R}.
\enq

The Coulomb law is often used to define sliding friction as follows:
\beq
\bm{f}=-\mu \frac{\bvp}{|\bvp|},\nn
\enq
where $\bvp$ is the slip velocity and $\mu$ is the coefficient of friction.
However, this definition of sliding friction is undefined at $\bvp=0$; 
therefore, it is difficult to analyze the equations of motion.
Thus, in this paper, to understand the dynamics of a rattleback with friction,
 I shall apply the viscous friction law  
 which states that friction is linearly related to $\bvp$ as follows:
\beq
\bm{f}=-\mu \bvp. \label{visf}
\enq

The angular momentum $\bm{L}$ has components in the principal inertia axes as follows
\beq
\bm{L} = \left(I_{10} \omega_1,I_{20} \omega_2,I_{30} \omega_3 \right),\nn
\enq
where $\omega_i$ are the components of the angular velocity $\bm{\omega}$ of the rattleback.
$\bm{\omega}$ is given by the equation $\ddt \bm{u}=0$.
In general, because the unit vectors of the principal inertia axes
${\bf e}_i$ have  time dependence, the time derivative of vector $\bm{A}$ is given by 
\beq
\frac{d}{dt} \bm{A}=\pdt \bm{A}+(\bm{\omega}\times \bm{A}),\nn
\enq
where $\pdt \bm{A}\equiv (\frac{d}{dt}A_i){\bf e}_i$.
Then
\beq
\bom = \pdt \bm{u} \times \bm{u} +n \bm{u}, \label{omega}
\enq
where $n \equiv \bom \cdot \bm{u}$.
The dynamical variables of these basic equations of motion are the components of 
the angular velocity $\omega_i$
and the slip velocity $v_{pi}$.

The equations of motion are obtained 
by dividing both sides of Eqs. (\ref{Ldot}) and (\ref{vgdot}) 
by $g$ and rescaling of the variables as follows:
\beq
\tilde{t} = \sqrt{g} t,\quad
\tilde{\bm{L}} = \frac{\bm{L}}{\sqrt{g}},\quad
\tilde{\bm{\omega}}=\frac{\bm{\omega}}{\sqrt{g}}, \nr
\tilde{R}=\frac{R}{g},\quad
\tilde{\bm{v}}_p=\frac{\bvp}{\sqrt{g}},\quad
\tilde{\mu}=\sqrt{g}{\mu}.\nn
\enq
In addition, the length of $c$ is assumed to be  unit length, i.e., $c=1$.

When a rattleback is turned lightly by hand such that the initial value of rotation 
is $\sim 2 \pi $ rad/s, spin reversal  occurs.
Because $g\sim 980 cm/s^2$, the spin value is $\tilde{n}\sim 0.201$.
As spin reversal occurs for this spin value or lower spin values, 
I  assume that $\tilde{n}=\frac{n}{\sqrt{g}}\sim O(10^{-2})$. 
This assumption is  the same as that for the neighborhood of the position
of stable equilibrium discussed by Pascal\cite{pas}. 
The value of the order of $\tilde{n}\sim O(10^{-2})$ 
is used in the next section to approximate the equations of motion.

In the case of limiting $\mu$ to infinity, these equations of motion lead to the no-slip case,
as shown by the numerical simulation performed in section \ref{sim}.
Therefore, to consider a case in which spin reversal occurs several times,
$\mu$ is assumed to be not as small as 
$\tilde{\mu}\sim O(10^2) \sim \frac{1}{\tilde{n}}$.

In the remaining sections, the tilde symbol above the variables is omitted.

%
%
\section{Effective equations of motion}\label{effecteq}
%
%

In this section, by applying the same linearization method as that of  Walker\cite{gtwalker}
 and Pascal\cite{pas},
 the effective equations of motion for a rattleback are obtained.
I set $c=1$ and $n\sim \frac{1}{\mu} \sim O(10^{-2})$, as discussed in the previous section.

The contact point $\bm{x}_p$ has components $(x,y,z)$,
 which satisfy $\frac{|x|}{a}<1,\frac{|y|}{b}<1$.
The second-order terms can be neglected.
Assuming that $\delta \sim O(10^{-2})$, $a \sim O(10)$, and $b \sim O(1)$,
as assumed in the numerical simulations,
Eqs. (\ref{p}), (\ref{q}), and (\ref{s}) for the parameters $p$, $q$, and $s$
are, respectively, approximated as
\beq
p\simeq \frac{1}{a^2},\quad
q\simeq \frac{\delta}{b^2},\quad
s\simeq \frac{1}{b^2}.\nn
\enq
These equations imply that the second-order terms of 
$\sqrt{p}x$, $\sqrt{s}y$, $\sqrt{q}x$, and $\sqrt{q}y$ can be neglected.
In addition, the order of the spin value $n\sim O(10^{-2})$ is used,
 as discussed in the previous section.
The second-order terms of $n$, $p$, $s$, and $q$ are also neglected.

For the orders of $\dot{x}$ and $\dot{y}$, 
when the time derivative of $x$ is considered,
$x$ is multiplied by the oscillation factor. 
Because a rattleback, unlike a usual top,
does not oscillate rapidly, $x$ and $\dot{x}$ are assumed to have the same order.
 In fact, as Eqs. (\ref{eqx}) and (\ref{eqy}) will show,
the oscillations are almost $\cos (\nu_{1,2} t)$, and according to  Eqs. (\ref{nu1}) and (\ref{nu2}),
$\nu_1^2 \sim \frac{5 a^2}{a^2+6}$ and $\nu_2^2 \sim \frac{5 b^2}{b^2+6}$, respectively.
When $a=10$ and $b=3$, as assumed in the simulation in section \ref{sim},
 $\nu_{1,2}$ are  $\nu_1^2 \sim 5$ and $\nu_2^2 \sim 3$; thus, it is assumed that
$\nu_{1,2} \sim O(1)$.
Therefore, the second-order terms of $O(\dot{x})\sim O(x)$,
and $O(\dot{y})\sim O(y)$, such as $psx \dot{y}$ are neglected.

In this approximation, by using Eqs. (\ref{xp}), (\ref{u}) and (\ref{omega}), 
$\bom = (\omega_1,\omega_2,\omega_3)$ has the following components,
\beq
\bom \simeq (q \dot{x}+s \dot{y},-p \dot{x}-q \dot{y},-n), \label{apo}
\enq
and $\bv_0 = \bm{x}_p \times \bom$ has components
\beq
\bv_0 \simeq(p \dot{x}+q \dot{y}-ny,
q \dot{x}+s \dot{y}+nx,0). \label{v0}
\enq
Because the second term of the time derivative of $\bm{v_0}$, $\bm{\omega}\times \bm{v_0} $,
is second order of $n$, $p$, $q$ and $s$, $\frac{d}{dt}\bm{v_0}$ are as follows:
\beq
\ddt\bm{v_0}  &\simeq& \pdt \bm{v_0}\nr
&\simeq& (p\ddot{x}+q \ddot{y}-n\dot{y},q \ddot{x}+s \ddot{y}+n\dot{x},0), \label{v0dotsim}
\enq
where terms such as $\dot{n}y$ are neglected because $\dot{n}$ is the second-order terms,
as will be shown in Eq. (\ref{I3dotn}).

In Eq. (\ref{R}) for R,
it is seen from Eqs. (\ref{u}) and (\ref{v0dotsim}) that $\left(\ddt \bm{v}_0\right)\cdot \bm{u}$
  is the second-order term; thus, $R \sim m $.
Therefore, setting $R=m$ in  Eqs. (\ref{Ldot}) and (\ref{vgdot}), the following equations are
obtained
\beq
\ddt \bm{L} &=& m (\bm{N}_0+\bm{N}_g+\bm{N}_p), \label{ldotn} \\
\ddt \bm{v}_g &=& \bm{f}, \label{vgf}
\enq
where 
\beq
\bm{N}_0 \equiv  \bm{x}_p \times \ddt\bv_0,\quad
\bm{N}_g \equiv  \bm{x}_p \times \bm{u},\quad
\bm{N}_p \equiv  \bm{x}_p \times \ddt\bv_p.\nn 
\enq
These equations contain six dynamical variables: $x$, $y$, $n$, and the components of $\bm{v}_p$.
Therefore, it is difficult to analyze of dynamics of these variables.

Using the approximation in Eq. (\ref{apo}) for $\bom$,
 and neglecting the second-order terms of $n$, $p$, $q$ and $s$,
the approximation of $\ddt \bm{L}$ has components
\beq
\ddt \bm{L} \simeq (I_1(q \ddot{x}+s\ddot{y}),-I_2(p\ddot{x}+q \ddot{y}),-I_3\dot{n}). \label{ld}
\enq
At the leading order, $\bm{N}_0$ has components
\beq
N_{01} &\simeq& -(q\ddot{x}+s\ddot{y})-n\dot{x},\nr
N_{02} &\simeq& p\ddot{x}+q\ddot{y}-n \dot{y}, \label{n0} \\
N_{03} &\simeq& q (x\ddot{x}-y\ddot{y})+sx\ddot{y}-p\ddot{x}y+n(x\dot{x}+y\dot{y}),\nn
\enq
where the approximation $n(1-p)x \dot{x} \sim n x\dot{x}$ is adopted.
At the leading order, $\bm{N}_g$ has components
\beq
\bm{N}_g \simeq (-y,x,q(y^2-x^2)+xy(p-s)), \label{ng}
\enq
where  the approximation $qx-(1-s)y \sim -y$ is adopted.

Next, I consider an approximation in which $\bm{v}_p$ is expressed by $x$, $y$, and $n$.
This approximation is crucial for simplifying these complicated equations of motion.
Eqs. (\ref{vgvp}), (\ref{visf}) and (\ref{vgf}) give $\mu \bvp=-\ddt\bv_0-\ddt\bvp$.
In addition, the numerical simulation in section \ref{sim} shows that $\ddt{\bv_0}\gg\ddt{\bvp}$;
therefore, the following approximation is obtained:
\beq
\bvp \sim -\frac{1}{\mu}\ddt \bv_0. \label{bvpsim}
\enq
Because the components of $\bv_0$ are given by $x$, $y$, and $n$ from Eq. (\ref{v0}),
the dynamical variables  reduce to $x$, $y$,
 and $n$ as will be shown in Eqs. (\ref{J1doty2})-(\ref{I3dotn2}).
It is easily seen from Eqs. (\ref{v0dotsim}) and (\ref{bvpsim}) that $v_{p3}$
is second-order term.
In addition, the term $\bom \times \bv_p$ is neglected in the time derivative of
$\bv_p$.
Thus, $\ddt\bv_p$ has components,
\beq
\ddt \bv_p \sim \pdt \bv_p \sim (\dot{v}_{p1},\dot{v}_{p2},0). \label{vpdot}
\enq
Using Eq. (\ref{vpdot}), $\bm{N}_p$ has components
\beq
\bm{N}_p \simeq (- \dot{v}_{p2}, \dot{v}_{p1},x\dot{v}_{p2}-y\dot{v}_{p1}). \label{np}
\enq
Substituting Eqs. (\ref{ld})-(\ref{np}) into Eq. (\ref{ldotn}) gives
\beq
J_1(q \ddot{x}+s \ddot{y})+n \dot{x} + y+\dot{v_{p2}}&=& 0,\label{J1doty}\\
J_2(p\ddot{x}+q\ddot{y})-n \dot{y}+ x +\dot{v_{p1}}&=& 0, \label{J2dotx}
\enq
where 
\beq
J_1\equiv I_1+1= \frac{b^2+6}{5},\quad
J_2\equiv I_2+1= \frac{a^2+6}{5}.\label{j1j2}
\enq
For $\dot{n}$, the following equation is obtained
\beq
&&I_3 \dot{n}+q (x \ddot{x}-y \ddot{y})+s x \ddot{y}-p \ddot{x}y +n(x\dot{x}+y\dot{y})\nn \\
&&+(q(y^2-x^2)+(p-s)xy)+x \dot{v_{p2}}-y \dot{v_{p1}} = 0. \label{I3dotn}
\enq

Let  $v_{p1},v_{p2}$ be expressed with respect to $x$, $y$ and $n$ from Eq. (\ref{bvpsim}).
Using Eqs. (\ref{v0dotsim}) and (\ref{J2dotx}) and adopting approximations 
$(1-\frac{1}{J_2}) \sim 1$ and $\frac{\dot{v_{p1}}}{\mu}\sim O(\frac{1}{\mu^2}) \sim 0$
 from Eq. (\ref{bvpsim}),
$v_{p1}$ is given by
\beq
v_{p1} &\simeq& -\frac{1}{\mu}(p \ddot{x}+q \ddot{y}-n \dot{y}), \nr
&=& -\frac{1}{\mu}\left(\frac{1}{J_2}(n\dot{y}-x-\dot{v_{p1}})-n\dot{y} \right),\nr
&\simeq&\frac{1}{\mu}(n \dot{y}+\frac{x}{J_2}). \label{vp1}
\enq
Similarly, $v_{p2}$ is given by
\beq
v_{p2} \simeq -\frac{1}{\mu}(n \dot{x}-\frac{y}{J_1}). \label{vp2}
\enq

By substituting Eqs. (\ref{vp1}) and (\ref{vp2}) in
Eqs. (\ref{J1doty})-(\ref{I3dotn}), and neglecting $O(\frac{n}{\mu})(<O(q))$ terms, 
such as $(J_1 q-\frac{n}{\mu}) \ddot{x} \sim J_1 q \ddot{x}$, at the leading order,
the following equations are obtained:
\beq
&&J_1(q \ddot{x}+s \ddot{y})+n \dot{x} + y+\frac{\dot{y}}{J_1 \mu}= 0,\label{J1doty2}\\
&&J_2(p\ddot{x}+q\ddot{y})-n \dot{y}+ x +\frac{\dot{x}}{J_2 \mu}= 0, \label{J2dotx2}\\
&&I_3 \dot{n}+q (x \ddot{x}-y \ddot{y})+s x \ddot{y}-p \ddot{x}y +n(x\dot{x}+y\dot{y})\nn \\
&&+(q(y^2-x^2)+(p-s)xy)+\frac{x \dot{y}}{\mu J_1}-\frac{y \dot{x}}{\mu J_2} = 0. \label{I3dotn2}
\enq

From the above expressions, it is observed that the main parts are oscillations of $x$ and $y$,
 such as
$\ddot{y}+\frac{1}{J_1 s}y=0$ and
$\ddot{x}+\frac{1}{J_2 p}x=0$.
Moffatt and Tokieda\cite{mt} suggested that
the terms $J_1 q \ddot{x}$, $n\dot{x}$, $J_2 q\ddot{y}$ and $n\dot{y}$
are crucial in creating reverse oscillations.
The effect of friction is included in the terms 
$\frac{\dot{y}}{J_1 \mu}$ and $\frac{\dot{x}}{J_2 \mu}$.

Furthermore, to eliminate terms depending on $q\ddot{x}$ and $q\ddot{y}$,
the new variables $X$ and $Y$ are defined as follows:
\begin{eqnarray}
\left(
\begin{array}{c}
X\\
Y\\
\end{array}
\right)
&\equiv&
T J P\left(
\begin{array}{c}
x\\
y\\
\end{array}
\right),
\label{xX}
\end{eqnarray}
where
\beqar
J \equiv \left(
\bear{cc}
\sqrt{J_2}&0\\
0&\sqrt{J_1}
\edar
\right),
P \equiv \left(
\bear{cc}
p&q\\
q&s\\
\edar
\right).\nn
\enqar 
In Eq. (\ref{xX}),  matrix $T$ is a rotational matrix that diagonalizes the symmetric matrix 
$Q\equiv  J^{-1} P^{-1}J^{-1} $, and it is defined by
\beq
T
&\equiv&\left(
\bear{cc}
\cos \theta&\sin \theta\\
-\sin \theta&\cos \theta
\edar
\right).\nn
\enq
The matrix  $Q$ is expressed as
\beq
Q
&=& \frac{1}{\Delta}\left(
\bear{cc}
\frac{s}{J_2}&-\frac{q}{\sqrt{J_1 J_2}}\\
-\frac{q}{\sqrt{J_1 J_2}}&\frac{p}{J_1}
\edar
\right).\nn
\enq
The rotational angle $\theta$ is given by
\beq
\tan 2\theta &=& -\frac{2 q \sqrt{J_1 J_2}}{J_1 s -J_2 p}\nr
&\sim& -\frac{2\sqrt{2}}{5}\delta a b \sim O(10^{-1})\label{tanth}
\enq
The eigenvalues of $Q$, $\nu_1$ and $\nu_2$,
correspond to the frequencies of $X$ and $Y$, respectively,
 and they are given as follows:
\beq
F&\equiv&TQT^{-1} =
\left(
\bear{cc}
\nu_1^2&0\\
0&\nu_2^2 \nn
\edar
\right).\nn
\enq
For $a>b$, by using Eqs. (\ref{p})-(\ref{s}) for the parameters 
$p$, $q$ and $s$, respectively,
eigenvalues $\nu_1$ and $\nu_2$ are given as follows:
\beq
\nu_1^2 
&\sim& 5\left(\frac{a^2}{a^2+6}+\delta^2 \frac{a^2}{6(1-\frac{a^2}{b^2})}\right),\label{nu1}\\
\nu_2^2 
&\sim& 5\left(\frac{b^2}{b^2+6}-\delta^2 \frac{a^2}{6(1-\frac{a^2}{b^2})}\right).\label{nu2}
\enq
The equations of motion with respect to  $X$ and $Y$ are given by
\beq
\left(D^2+F+DRF\right)\left(\bear{c}
X\\
Y\\ \end{array} \right)
=\left(\bear{c} 0\\0\\ \edar\right) \nn
\enq
where
\beq
R &\equiv&TJ^{-1}NJT^{-1},\nr
N&\equiv& \left( \bear{cc} 0&-n \\ n&0 \\ \edar \right),
D \equiv \left( \bear{cc} \ddt&0 \\0&\ddt\\ \edar \right).\nn
\enq
The matrix $R$ has components $a_i$ as follows:
\beq
R&=& \left( \bear{cc} a_1&a_2 \\a_3&a_4 \\ \edar \right),
\enq
where $a_i$ are defined by
\beq
&&a_1 \equiv m_1-n k_1,\, 
a_2 \equiv -m_2-n k_2,\, \nr
&&a_3 \equiv -m_3+n k_3,\, 
a_4 \equiv m_4+n k_4.\label{ai}
\enq
In these equations, $k_i$ are given by
\beq
k_1 &\equiv& -\nu_1^2 \sin \theta \cos \theta(f-f^{-1}) 
\sim a^2 \delta \sim O(1),\label{k1} \\
k_2 &\equiv& \nu_2^2 (f\sin^2 \theta+f^{-1}\cos^2 \theta) 
\sim \frac{5b}{\sqrt{2}a}\sim O(1), \\
k_3 &\equiv& \nu_1^2 (f\cos^2 \theta+f^{-1}\sin^2 \theta)
\sim \frac{5a}{\sqrt{2}b} \sim O(10), \\
k_4 &\equiv& -\nu_2^2 \sin \theta \cos \theta(f-f^{-1})
\sim \frac{\delta a^2}{2} \sim O(1),\label{k4}
\enq
and $m_i$ depending on friction parameter $\mu$ are given as follows:
\beq
m_1 &\equiv& \frac{f}{J_2 \mu}\frac{\nu_1^2}{\nu_2^2} k_2,\quad
m_2 \equiv \frac{f}{J_2 \mu} k_4, \label{m12}\\
m_3 &\equiv& \frac{f}{J_2 \mu} k_1, \quad
m_4 \equiv \frac{f}{J_2 \mu}\frac{\nu_2^2}{\nu_1^2} k_3. \label{m34}
\enq
where $f$ is defined by
\beq
f\equiv\sqrt{\frac{J_2}{J_1}}=\sqrt{\frac{a^2+6}{b^2+6}}. \label{f}
\enq

Finally, the effective equations of motion are obtained as follows:
\beqar
&&\ddot{X}+\nu_1^2 X  +a_1 \dot{X}+a_2 \dot{Y}  =0,\label{eqx}\\
&&\ddot{Y}+\nu_2^2 Y +a_3 \dot{X}+a_4 \dot{Y} =0,\label{eqy}\\
&&I_3 \dot{n} -(X,Y)
\left(KD^2-K-SD \right)\left(\bear{c} X \\ Y \edar \right) = 0.\label{eqn}
\enqar
Here, $S$ and $K$ are 2$\times$2 matrices as follows:
\beqar
&&K \equiv \left(\bear{cc} k_1&-k_3\\
k_2&-k_4 \edar \right),\quad
S\equiv \left(\bear{cc} s_1&s_2\\
s_3&s_4 \edar \right),
\enqar
where $s_i$ are defined by
\beq
s_1 &\equiv& \nu_1^2(n \sqrt{J_1J_2}k_3-\frac{k_1}{\mu}),\nr
s_2 &\equiv& \nu_2^2 (n \sqrt{J_1J_2}k_1+\frac{k_3}{\mu}),\nr
s_3 &\equiv& \nu_1^2 (n \sqrt{J_1J_2}k_4-\frac{k_2}{\mu}),\nr
s_4 &\equiv& \nu_2^2 (n \sqrt{J_1J_2}k_2+\frac{k_4}{\mu}).\nn
\enq

From Eqs. (\ref{m12}) and (\ref{m34}), it is found that these $m_i$ are approximated 
as $m_i \sim \frac{k_i}{a \mu}$.
When $n \sim O(10^{-2})$ and $\frac{1}{\mu} \sim O(10^{-2})$ are considered,
$a_i$ take a value of the order $O(10^{-1})$; therefore,
 I will neglect the quantities of $O(a_i^2)$ in the next section.
For the no-slip case - ($\mu$ is sufficiently large),
\beq
m_i\sim 0. \label{miap}
\enq
Thus, it is observed that the effect of friction is included 
as $\frac{1}{\mu}$ in the  $m_i$ parameters.
%
%
\section{Slowly varying mode}\label{svm}
%
%
 The motion of a rattleback contains a rapid frequency rattling mode  and 
 a slowly varying amplitude mode. To discuss the number of spin reversals,
 the slowly varying mode should be analyzed. 
 In this section,  by approximately solving the characteristic equation
  and considering the  time average of the rapid frequency mode,
  the  equations of the slowly varying mode is derived
  from the effective equations of motion which were obtained in the previous section.

To study the mode contained in $X(t)$ and $Y(t)$,
I approximately solve the characteristic equation given in Eqs. (\ref{eqx}) and (\ref{eqy}).
In this case, $n$ is considered to be a constant, 
because $\dot{n}$ is a second-order term from Eq. (\ref{I3dotn2}).
Moreover, the quantities of $O(a_i^2)$ are neglected as explained in the previous section.

Two modes are observed, which correspond to the rattling motion 
of the long axial direction $e^{i \nu_1 t} e^{-\frac{a_1}{2}t}$
 and the short axial direction $e^{i \nu_2 t} e^{-\frac{a_4}{2}t}$.
Then,  the mode expansions of $X(t)$ and $Y(t)$ are given by
\beq
X(t) &=& c_1 e^{i\lambda_1 t} +c_1^* e^{-i \lambda_1^* t}
+c_2 e^{i\lambda_2 t} +c_2^* e^{-i \lambda_2^* t}, \nr 
Y(t) &=& d_1 e^{i\lambda_1 t} +d_1^* e^{-i \lambda_1^* t}
+d_2 e^{i\lambda_2 t} +d_2^* e^{-i \lambda_2^* t},\nn
\enq
where $\lambda_1 \equiv \nu_1+i \frac{a_1}{2},\lambda_2 \equiv \nu_2+i \frac{a_4}{2}$, and 
$*$ indicates a complex conjugate. 
The coefficients
$c_1$, $c_2$, $d_1$ and $d_2$ are obtained from $X(t)$ satisfying the equation of motion (Eq. (\ref{eqx}))
 and initial conditions such as $X(0)=x_0$, $Y(0)=y_0$, $\dot{X}(0)=0$ and $\dot{Y}(0)=0$ as follows:
\beq
c_1 &\sim&\left(\frac{1}{2}-i\frac{a_1}{4 \nu_1}\right)x_0+i\frac{a_2 \nu_2^2}{2\nu_1 \Delta} y_0, \nr
c_2 &\sim& -i \frac{a_2 \nu_2}{2\Delta}y_0, \nr
d_1 &\sim&-i \frac{a_1^2}{8 a_2 \nu_1} x_0, \nr
d_2 &\sim& -i \frac{a_1^2}{8 a_2 \nu_2}x_0+\left(\frac{1}{2}-i\frac{a_4}{4 \nu_2}\right)y_0,\nn
\enq
where $\Delta \equiv \nu_1^2-\nu_2^2$.

To obtain the behavior of $n(t)$,
I consider the time average of the terms in the rapidly varying mode  such as $e^{i \nu_i t}$, 
which are contained in terms, such as $X^2$ and $X\dot{X}$, of the equation for $\dot{n}(t)$.
When the variables  are given by $Q_1(t)=F_1(t)S_1(t)$ and $Q_2(t)=F_2(t)S_2(t)$,
(where $F_i(t)$ are rapidly varying functions and $S_i(t)$ are slowly varying functions
such as $e^{-\frac{a_i}{2}t}$),
the time average of $Q_1$ and $Q_2$ ($\langle Q_1Q_2\rangle$) is defined as follows:
\beq
\langle Q_1Q_2\rangle \equiv S_1S_2\lim_{T \rightarrow \infty }\frac{1}{T}\int_0^T F_1F_2 dt.\nn
\enq
Then, neglecting terms $O(a_i^2)$,
\beq
\langle X^2\rangle &\sim& \frac{x_0^2}{2} e^{-a_1 t},\quad
\langle X\dot{X}\rangle \sim -\frac{x_0^2}{4} a_1 e^{-a_1 t,} \nr
\langle X \ddot{X}\rangle &\sim& -\frac{\nu_1^2 x_0^2}{2}e^{-a_1 t}, \nr
\langle X\dot{Y}\rangle &\sim& \frac{a_1^2}{8 a_2} x_0^2 e^{-a_1 t}
-\frac{a_2 \nu_2^2}{2 \Delta}y_0^2e^{-a_4 t}
\sim -\langle Y \dot{X}\rangle,\nr
\langle Y^2\rangle &\sim& \frac{y_0^2}{2} e^{-a_4 t},\quad \label{YY}
\langle Y\dot{Y}\rangle \sim -\frac{y_0^2}{4} a_4 e^{-a_4 t}, \nr
\langle Y \ddot{Y}\rangle &\sim& -\frac{\nu_2^2 y_0^2}{2}e^{-a_4 t}, \nr
\langle XY\rangle &\sim& 0, \quad \langle X \ddot{Y}\rangle \sim 0.\nn
\enq
From these equations, it is observed that $\langle X\dot{X}\rangle$ and $\langle Y\dot{Y}\rangle$
are $O(a_i)$.
Thus, the main part of $\dot{n}$ is obtained as follows:
\beq
I_3 \dot{n}(t) 
&\sim& k_1(\langle X \ddot{X}\rangle-\langle X^2\rangle)
-k_4(\langle Y \ddot{Y}\rangle-\langle Y^2\rangle)\nn \\
&=& -\frac{k_1}{2} A^2 +\frac{k_4}{2} B^2, \label{ndotab}
\enq
where
\beq
A&\equiv& \sqrt{\nu_1^2-1}x_0 e^{-\frac{a_1}{2}t},
B\equiv \sqrt{\nu_2^2-1}x_0 e^{-\frac{a_4}{2}t}. \label{abdef}
\enq
From the definitions given in Eq. (\ref{abdef}), these variables satisfy the differential equations 
by using the approximation $\dot{n} \sim 0$ as follows:
\beq
\dot{A} = -\frac{a_1}{2} A, \quad \dot{B} = -\frac{a_4}{2} B. \label{abdot}
\enq
In the no-slip case ($\mu$ is sufficiently large), from Eq. (\ref{miap}), 
the parameters $a_1$ and $a_4$ are given as $a_1=-nk_1$ and $a_4=nk_4$, respectively.
Then, substituting these parameters into Eqs. (\ref{ndotab}) and
(\ref{abdot}) gives  equations that correspond to those obtained by 
Pascal\cite{pas}, Markeev\cite{mar} and Moffatt and Tokieda\cite{mt}.

Multiplying Eq. (\ref{ndotab}) by $n$ and using  Eqs. (\ref{ai}) and (\ref{abdot}),
the following equation is obtained:
\beq
\ddt(N^2+ A^2+ B^2)=-m_1  A^2-m_4  B^2, \label{slipnab}
\enq
where $N\equiv \sqrt{I_3}n$.
For no-slip case ($m_i=0$), the variable
$E_1\equiv N^2+ A^2+ B^2$  corresponds to the sum of energy
about the rotation $N$, and the amplitudes of the long and short axial directions
( $A$ and $B$, respectively) are conserved.
Moreover, from Eq. (\ref{abdot}), it is observed that 
the variable $E_2 \equiv A^\gamma B$ is conserved, where $\gamma \equiv \frac{k_4}{k_1}$.
In the phase spaces of $N$, $A$ and $B$, $E_1$ corresponds to a sphere and $E_2$ corresponds 
to a quasi-hyperbolic cylinder. The trajectories of the system are closed curves that intersect
this cylindrical surface and the sphere. 
Therefore, an infinite number of spin reversals is obtained.

In contrast, for the slip case, 
it is observed that the energy $E_1$ decreases according to the right-hand-side of Eq. (\ref{slipnab}),
which relates to friction. By using the approximation $\dot{n} \sim 0$,
it is assumed that $\gamma$ is almost constant and $E_2$ is almost conserved; 
then, the trajectory of the system decreases as
the radius of the sphere of $E_1$ decreases. Therefore, a finite number of spin reversals is obtained.

%
%
\section{New phenomenon for the slip case} \label{newph}
%
%
For the no-slip case,  rattle vibration increases irrespective of how small the initial
spin  $n_0$ is, and in theory, spin reversal occurs.
However, for the slip case, when $n_0$ is small, vibration does not increase,
 and rotation stops. Even if  rattle vibration occurs, spin reversal does not.
Thus, a critical value of the spin  $n_{c1}$ may exist under which  rattle vibration does not
increase, and a critical value of the spin  $n_{c2}$ may exist under which  rattle
vibration increases but spin reversal does not occur.
Therefore, it is assumed that  the number of spin reversals is finite 
due to the existence of these  critical spin values.
In this section, I discuss some facts relating to these  values.
%
\subsection{Critical spin  $n_{c1}$ necessary to increase  rattle vibration} \label{nc1}
%

For spin reversal to occur, rattle vibration must increase.
This vibration increases according to the factor $e^{-a_1 t}$ and $e^{-a_4 t}$.
For the no-slip case, these factors are $e^{+n k_1 t}$ and $e^{-n k_4 t}$.
From Eqs. (\ref{k1}) and (\ref{k4}), it is observed that
$n k_1$ and $n k_4$ have the same sign and if one mode increases,
the other decreases. 
In contrast, for the slip case, $m_1$ and $m_4$ are in $a_1$ and $a_4$;
hence, $a_1>0$ and $a_4>0$  occur according to the value of $n$.
In this case, both modes decrease, and  rattle vibration does not increase.
The spin values $n_{c1\pm}$ are given by
\beq
n_{c1+}&\equiv& \frac{m_1}{k_1}
=-\frac{\cos^2 \theta+f^2 \sin^2 \theta}{J_2 \mu \sin\theta \cos \theta
(f-f^{-1})}, \nr
n_{c1-}&\equiv& -\frac{m_4}{k_4} 
= \frac{f^2 \cos^2 \theta+ \sin^2 \theta}{J_2 \mu \sin\theta \cos \theta
(f-f^{-1})}, \nn
\enq
for  $k_1>0$.
For the case of $k_1<0$, these values reverse and are given by 
$n_{c1+}= -\frac{m_4}{k_4}$ and $n_{c1-}=\frac{m_1}{k_1}$.
Therefore, it is observed that  rattle vibration decreases
for the initial spin $n_0$ in the range $n_{c1-}<n_0<n_{c1+}$.
 
%
\subsection{Critical spin  $n_{c2}$ necessary to reverse rattleback spin} \label{nc2}
%
When the spin is more than $n_{c1}$, rattle vibration increases.
However, the spin does not necessarily reverse. 
A critical spin value $n_{c2}$ may exist over which spin reversal occurs.

When starting with $n_0>0$, although it is near the value of $n_0$,
for a while, rattle vibration begins rapidly, and  spin decreases. 
Eventually, the rattleback stops spinning and then reverses direction,
and the spin value decreases to $n_1<0$. 
This spin value  $n_1$ is obtained as follows. 

Now,  consider the case for $k_1>0$, $k_4>0$ and $n_0>0$.
At first, the modes $A$ and $B$  exponentially increase and decrease,
respectively: therefore, from Eq. (\ref{ndotab}),
$\dot{n}$ approximates to
\beq
I_3 \dot{n} \simeq -\frac{k_1 \nu_1^2}{2} A^2. \label{ndota2}
\enq
Integrating this equation from $t_0(n=n_0>0)$ to $t_1(n=n_1)$ gives 
\beq
I_3 (n_1-n_0) = -\frac{k_1 \nu_1^2}{2} \int_{t_0}^{t_1} A^2. \label{n1n0}
\enq
Furthermore, Eq. (\ref{slipnab}) approximates to
\beq
\ddt(I_3 n^2+\nu_1^2 A^2)\simeq-m_1 \nu_1^2 A^2. \label{slipnab2}
\enq
Integrating this equation gives
\beq
\nu_1^2 \int_{t_0}^{t_1} A^2 \simeq -\frac{I_3}{m_1}(n_1^2-n_0^2), \label{a2int}
\enq
where it is  considered that $A(t_0)\sim 0$ and $A(t_1) \sim 0$.
By substituting Eq. (\ref{a2int}) into Eq. (\ref{n1n0}), $n_1$ is given by
\beq
n_1 = -n_0+\frac{2 m_1}{k_1}.
\enq
Then it is observed that if $n_1< 0$, that is, the initial spin $n_0> \frac{2 m_1}{k_1}$,
 one spin reversal occurs,
and if $n_0 \leq \frac{2 m_1}{k_1}$, no spin reversal occurs.
Therefore, it is observed that if $n_{c1+}=\frac{m_1}{k_1}<n_0\leq\frac{2 m_1}{k_1}$,
rattle vibration increases but spin reversal does not occur.
Similarly, when $n_0<0$, the condition $n_0< -\frac{2 m_4}{k_4}$ must be satisfied
for spin reversal to occur. 

Finally, it is observed that there exist critical spins $n_{c2\pm}$ as follows:
\beq
n_{c2+}&\equiv& \frac{2 m_1}{k_1},\quad
n_{c2-}\equiv -\frac{2 m_4}{k_4}.\nn
\enq
When $n_{c2-}\leq n_0<n_{c1-}$ or $n_{c1+}<n_0\leq n_{c2+}$,
rattle vibration increases but  spin reversal does not occur.
For the case of $k_1<0$, it is observed that
$n_{c2-}$ and $n_{c2+}$ are switched:
$n_{c2-}= \frac{2 m_1}{k_1}$ and $n_{c2+}= -\frac{2 m_4}{k_4}$.

%
\subsection{Number of spin reversals $n_r$} \label{rn}
%
For the no-slip case, because dynamical energy is conserved,
an infinite number of spin reversals occur.
In contrast, for the slip case, because friction exist,
the number of spin reversals $n_r$ is finite.

As discussed in the previous section, for $n_1< 0$, the condition for the first spin reversal 
occurrence is given by 
\beq
\frac{1}{h_1}> 1, \label{h11}
\enq
 where 
\beq
h_1\equiv \left| \frac{2 m_1}{k_1 n_0} \right| .
\enq
Next, for the second spin reversal, if $n_1$ satisfy $n_1>n_{c1-}=-\frac{m_4}{k_4}$,
rattle vibration does not increase and the rattleback stops spinning. 
If $n_1$ satisfy $n_{c1-}>n_1>n_{c2-}=-\frac{2 m_4}{k_4}$, rattle vibration increases
but spin reversal does not occur. Therefore, the condition for
the second spin reversal to occur is as follows:
\beq
n_1<-\frac{2 m_4}{k_4}. \label{srcond}
\enq

Now, I shall estimate  spin $n_2$ after the second rattle vibration.
Similar to the derivation of $n_2$ in the previous section,
 the following equation is obtained:
\beq
n_2&=&-n_1-\frac{2 m_4}{k_4} \nr
&=&n_0-\frac{2 m_1}{k_1}-\frac{2 m_4}{k_4}.\nn
\enq
Thus, if $n_2> 0$, the second spin reversal occurs. 
This condition is consistent with Eq. (\ref{srcond}).
The condition for the second spin reversal occurrence is given by
\beq
\frac{1}{h_1}> 1+\rho, \label{h12}
\enq
where
\beq
\rho \equiv \frac{h_4}{h_1}, \quad h_4 \equiv \left|\frac{2 m_4}{k_4 n_0}\right|.\nn
\enq
From Eqs. (\ref{h11}) and (\ref{h12}), the condition that 
spin reversal occurs only once is given by $1< \frac{1}{h_1}\leq 1+\rho$.
Similarly, the condition that spin reversal occurs only two times
is given by $1+\rho< \frac{1}{h_1}\leq 2+\rho$, only three  is given by 
$2+\rho< \frac{1}{h_1}\leq 2+2\rho$, and so on.

It is observed that the number of spin reversals $n_r$ satisfies the 
following inequalities:

\vspace{0.5cm}
（I）$n_0>0,k_1>0,k_4>0$ or $n_0<0,k_1<0,k_4<0$ 
\beq
K_0(n_r-1,\rho)&<&\frac{1}{h_1}\leq K_0(n_r,\rho), \label{nrk0}
\enq

（II）$n_0>0,k_1<0,k_4<0$ or $n_0<0,k_1>0,k_4>0$
\beq
K_1(n_r-1,\rho)&<&\frac{1}{h_1}\leq K_1(n_r,\rho), \label{nrk1}
\enq
where 
\beq
K_{0}(n,\rho) &\equiv& \sum^{n}_{i=0} P_{+}(i)+P_{-}(i) \rho, \nr
K_{1}(n,\rho) &\equiv& \sum^{n}_{i=0} P_{+}(i) \rho+P_{-}(i),\nr
P_{\pm}(i) &\equiv& \frac{1\pm(-1)^i}{2}. \nn
\enq
Finally, from the inequalities stated in Eqs. (\ref{nrk0}) and (\ref{nrk1}), 
the number of spin reversals $n_r$ are obtained.

For case (I), $n_r$ is given by
\beq
n_r = \left\{
\bear{ll}
2[x]+1&\mbox{ if $x-1\leq[x]< x-\frac{1}{1+\rho}$}\\
2[x]&\mbox{ if $x-\frac{1}{1+\rho}\leq[x]< x$}
\edar
\right.
\label{nr1}
\enq
and for case  (II), $n_r$ is given by
\beq
n_r = \left\{
\bear{ll}
2[x]+1&\mbox{ if $x-1\leq[x]< x-1+\frac{1}{1+\rho}$}\\
2[x]&\mbox{ if $x-1+\frac{1}{1+\rho}\leq[x]< x$}
\edar
\right.
\label{nr2}
\enq
where $x$ is defined by
\beq
x \equiv \frac{1}{h_1(\rho+1)},\nn
\enq
and $[x]$ is Gauss' symbol, which represents the greatest integer 
 less than or equal to $x$.

%
\subsection{Relationship between  number of spin reversals and rattleback shape} \label{rn2}
%

In this subsection, the relationship among the number of spin reversals, the axis
 inclination ($\delta\sim \theta$ in Eq. (\ref{tanth})), the coefficient of friction $\mu$ and 
the rattleback shape are obtained.

The factor $x=\frac{1}{h_1(\rho+1)}$ is written as
\beq
x=\frac{1}{2}J_2\mu|n_0 \sin \theta \cos \theta| L(f), \nn
\enq
where $L(f)$ is the form factor given by
\beq
L(f) \equiv \frac{|f-f^{-1}|}{1+f^2}. \label{ff}
\enq
This expression shows that the number of spin reversals is proportional to 
the coefficient of friction $\mu$, the axis inclination $\delta$,
and the initial spin $n_0$.
Therefore, when the friction is large (close to the no-slip case)
 and the asymmetry between the shape axes and the principal inertia axes is large, 
the number of spin reversal becomes large.

\begin{figure}[h]
  \begin{center}
    \includegraphics[width=0.45\textwidth]{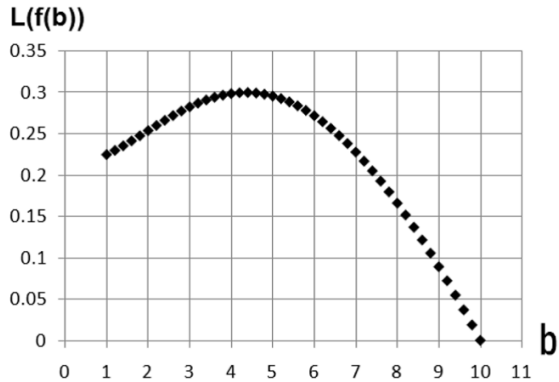}
    \caption{Form factor $L(f)$ versus $b$ for $a = 10$\label{lf}}
  \end{center}
\end{figure}

Here, I discuss the form factor $L(f)$.
When the long axis $a$ is fixed, which responds to a fixed $J_2$,
enlarging $f$ correspond to reduction in $J_1$ 
from Eq. (\ref{f}).
Furthermore, from Eq. (\ref{j1j2}), reducing $J_1$ corresponds to a reduction in $b$,
which indicates that the rattleback becomes long and slender.
In contrast, reducing $f$ results in the rattleback becoming more circular,
 such as $J_1=J_2$.
The form factor $L(f)$ has maximum at $f=\sqrt{2+\sqrt{5}}$.
Thus, it is observed that if the  rattleback  long and slender,
rattle vibration about the short axis does not  easily occur
and the number of spin reversals is reduced.
This also occurs, if the rattleback is more circular.
The relationship between $a$ and $b$ corresponding to the maximum of $f$ 
is given by
\beq
b^2 = \frac{a^2+6}{2+\sqrt{5}}-6.
\enq
For example, if $a=10$, $b$ becomes approximately 4.4.
Fig. \ref{lf} shows the change in the form factor $L(f)$ when
$a=10$ is fixed and $b$ is varied from 1 to 10.
It is observed that there is a maximum value around $b\sim 4.4$.

%
%
\section{Numerical results}\label{sim}
%
%
　In this section, the results of computations based on
the exact system described in Eqs. (\ref{Ldot})-(\ref{visf}) are presented
and these results are compared  with those based on the effective equations of motion
described in Eqs. (\ref{eqx})-(\ref{eqn}).

For the numerical simulations, the NDsolve command in $Mathematica$ is used.

\subsection{Validity of the approximation $\dot{\bm{v}}_0\gg \dot{\bvp}$}
In section \ref{effecteq}, the approximation $\dot{\bm{v}}_0\gg \dot{\bvp}$ was used
to derive effective equations of motion.

\begin{figure}[h]
\begin{minipage}{0.8\linewidth}
    \includegraphics[width=\linewidth]{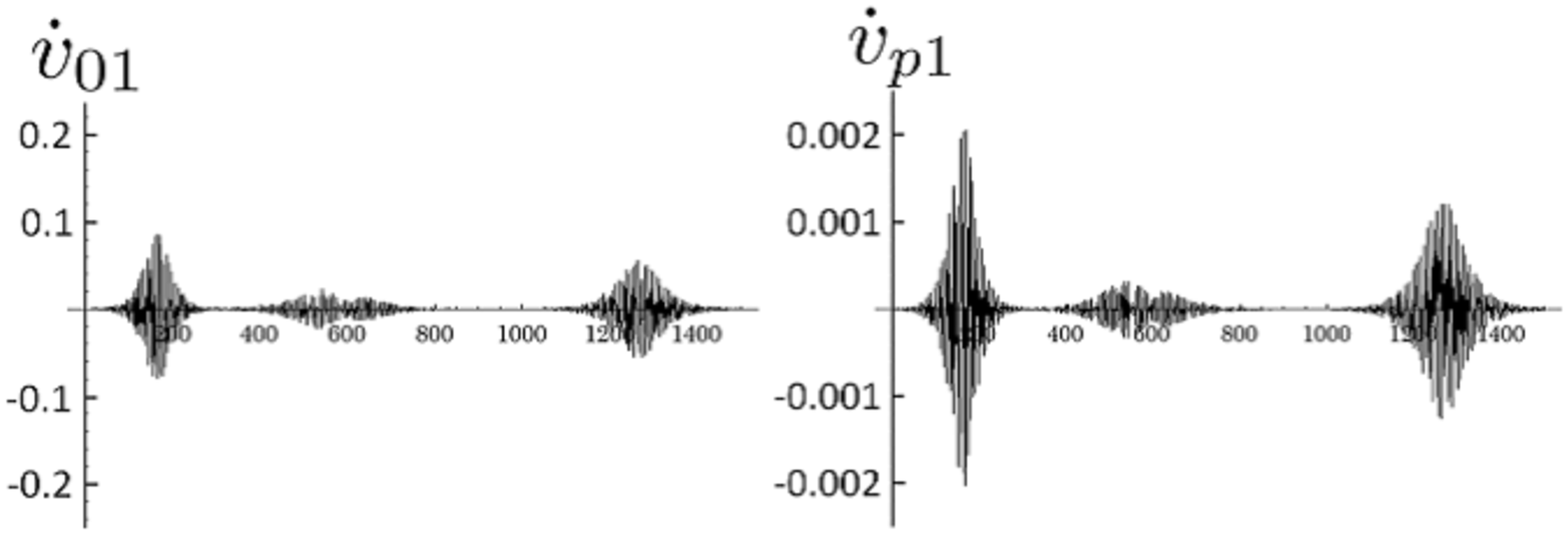}
    \caption{Time evolution of $\dot{v}_{01}$  and $\dot{v}_{p1}$
     for $\mu = 100$  \label{v01vp1m100}}
\end{minipage}
\begin{minipage}{0.8\linewidth}
    \includegraphics[width=\linewidth]{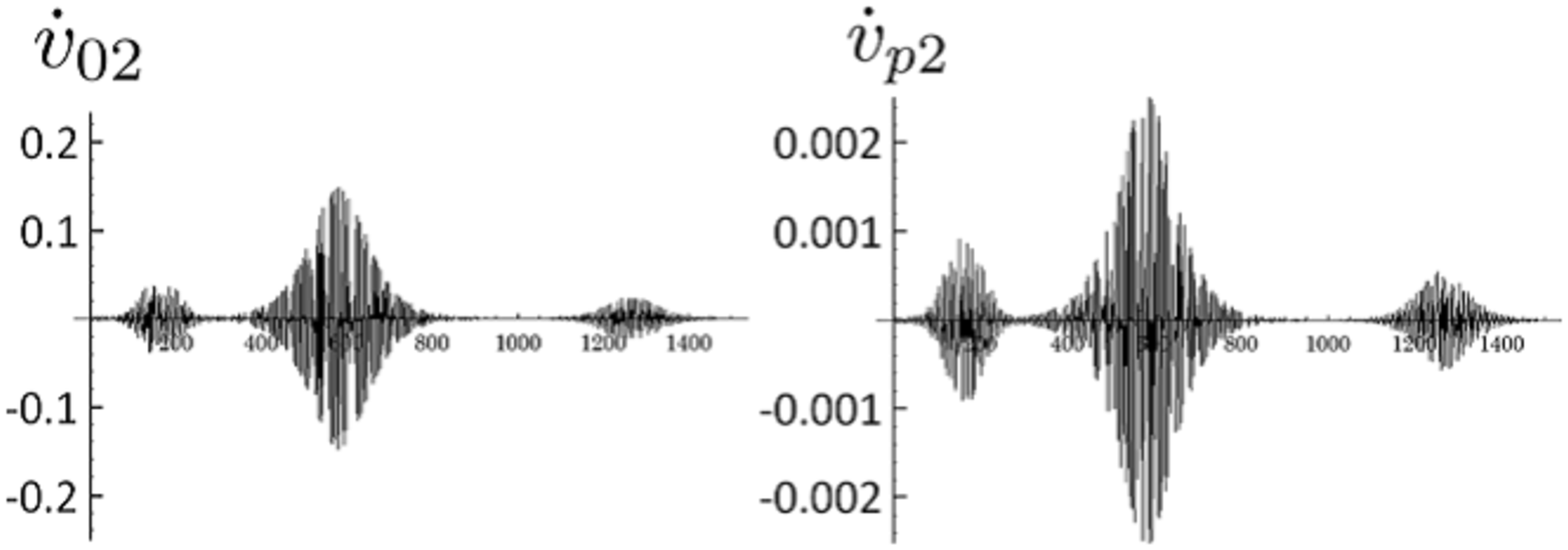}
    \caption{Time evolution of $\dot{v}_{02}$ and $\dot{v}_{p2}$ 
    for $\mu = 100$  \label{v02vp2m100}}
\end{minipage}
\end{figure}

\begin{figure}[h]
\begin{minipage}{0.8\linewidth}
    \includegraphics[width=\linewidth]{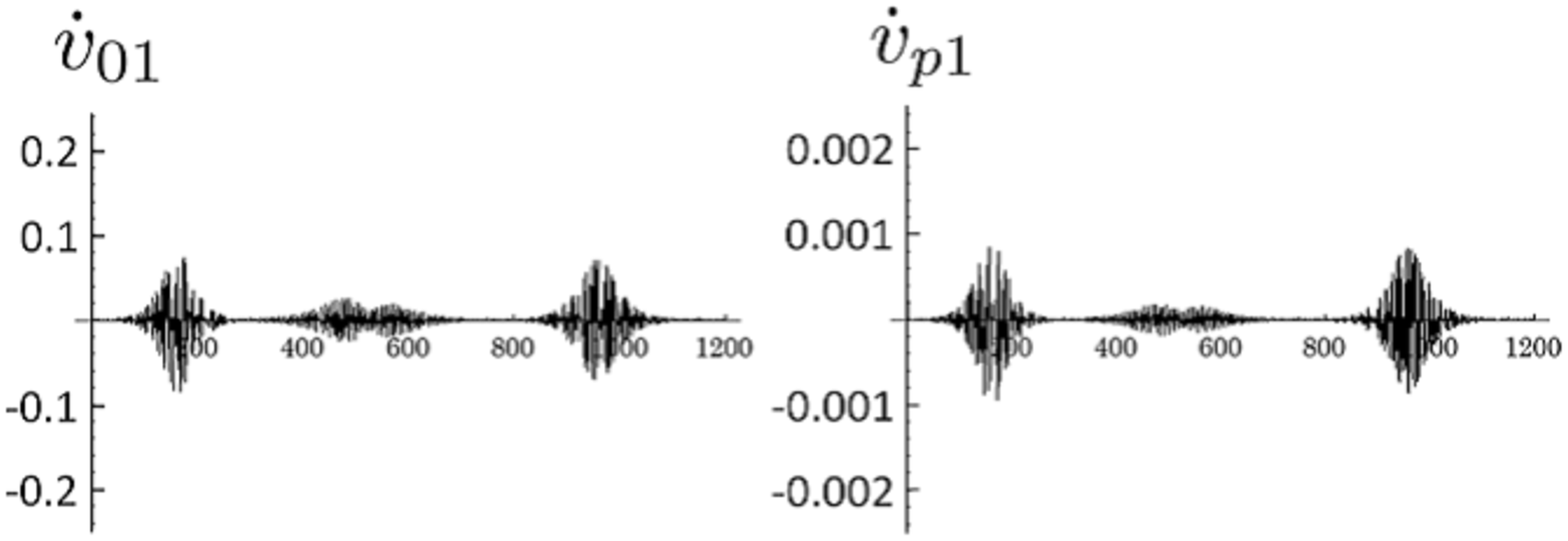}
    \caption{Time evolution of $\dot{v}_{01}$  and $\dot{v}_{p1}$
     for $\mu = 200$ \label{v01vp1m200}}
\end{minipage}
\begin{minipage}{0.8\linewidth}
    \includegraphics[width=\linewidth]{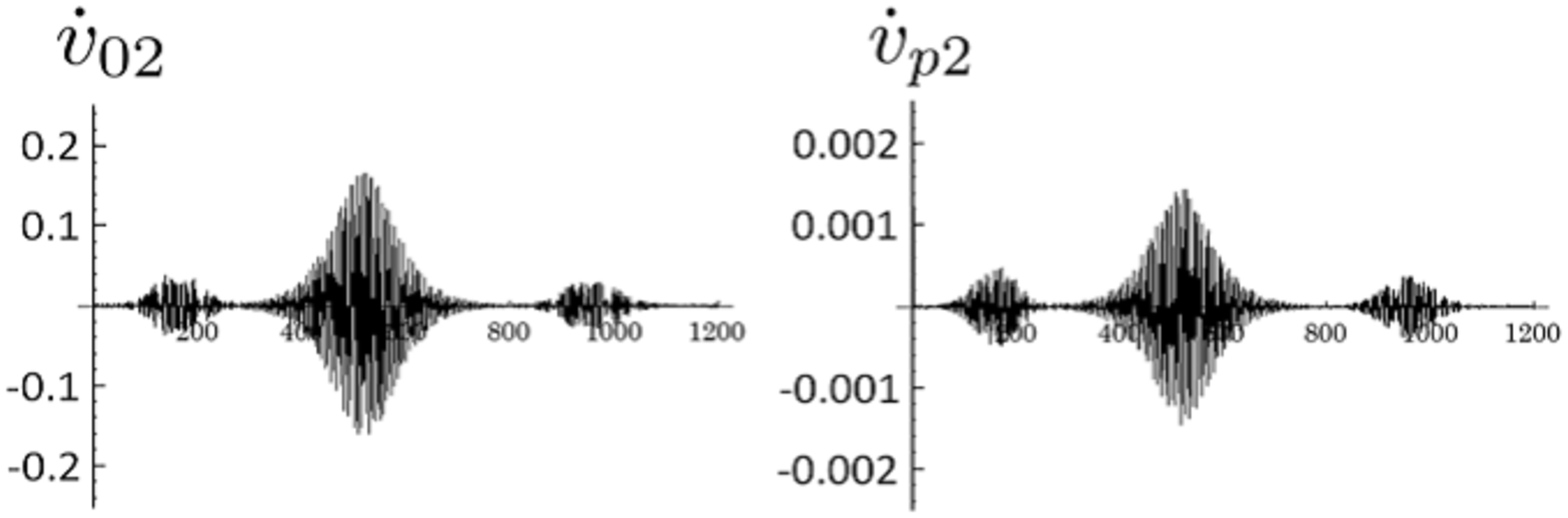}
    \caption{Time evolution of $\dot{v}_{02}$ and $\dot{v}_{p2}$ 
    for $\mu = 200$   \label{v02vp2m200}}
\end{minipage}
\end{figure}

The initial conditions are $x_0=y_0=0.01$, $\dot{x}_0=\dot{y}_0=0$ 
and $n_0=0.05$ with parameters $a=10$, $b=3$, and $\delta=0.03$.
Fig. \ref{v01vp1m100} shows the behavior of $\dot{v}_{01}$ 
and $\dot{v}_{p1}$ for $\mu = 100$, and Fig. \ref{v02vp2m100} 
shows the behavior of $\dot{v}_{02}$ and $\dot{v}_{p2}$ for $\mu = 100$.
From Figs. \ref{v01vp1m100} and \ref{v02vp2m100},
the comparison of the amplitudes shows that
$\dot{v}_{p1}\sim \dot{v}_{01}\times \frac{1}{50}$ and 
$\dot{v}_{p2}\sim \dot{v}_{02}\times \frac{1}{50}$; therefore, it can be safely
assumed that 
$\dot{v}_{01}\gg \dot{v}_{p1}$ and 
$\dot{v}_{02}\gg \dot{v}_{p2}$.
Figs. \ref{v01vp1m200} and \ref{v02vp2m200} are the same as 
Figs. \ref{v01vp1m100} and \ref{v02vp2m100}, respectively,
 but for $\mu = 200$.
In this case, it is similarly shown that
$\dot{v}_{p1}\sim \dot{v}_{01}\times \frac{1}{100}$ and 
$\dot{v}_{p2}\sim \dot{v}_{02}\times \frac{1}{100}$;
therefore, $\dot{v}_{01}\gg \dot{v}_{p1}$ and
$\dot{v}_{02}\gg \dot{v}_{p2}$.
Note that the approximation  improves when $\mu$ increases.
Thus, it can be concluded that the approximation $\dot{\bm{v}}_0\gg \dot{\bvp}$ is valid.

\subsection{$n_{c1\pm}$, $n_{c2\pm}$, and spin reversal behavior}
In subsections \ref{nc1} and \ref{nc2} of section \ref{newph}, the critical
values of spin necessary to increase rattle vibration and cause spin reversal are obtained.
Here, computational results for  these values and the spin reversal behavior 
which are  obtained from a simulation  based on
the exact system are presented.

\begin{figure}
   \includegraphics[width=0.35\textwidth]{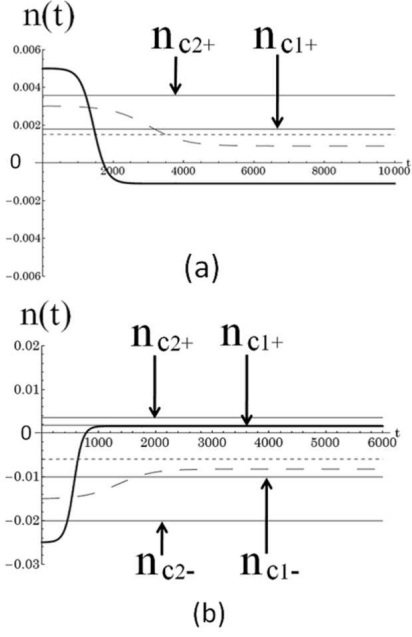}
    \caption{\small{Time evolution of $n(t)$ with (a) positive initial spin 
    with $n_0=0.0015$ (dotted line), $0.003$ (dashed line),
     and $0.005$ (solid line) and (b) negative initial spin with
    $n_0=-0.006$ (dotted line), $-0.015$ (dashed line) and $-0.025$ (solid line). }
  \label{nc21pm}}
\end{figure}

The initial conditions are $x_0=y_0=0.01$, and $\dot{x}_0=\dot{y}_0=0$ with 
parameters $a=10$, $b=3$, and $\delta=0.03$.
The coefficient of friction is set to  $\mu = 75$.
For these initial conditions, the following critical spin are obtained:
$n_{c2+}=0.00358$, $n_{c1+}=0.00179$, $n_{c1-}=-0.01$, and $n_{c2-}=-0.0201$.
Fig. \ref{nc21pm} shows the computed evolution of $n(t)$.
In  Fig. \ref{nc21pm} (a), as discussed in  subsection \ref{nc1}, 
the dotted line with the initial spin $n_0=0.0015<n_{c1+}$ shows that
 rattle vibration does not increase.
As discussed in  subsection \ref{nc2}, the dashed line with the initial spin
$n_{c1+}<n_0=0.003<n_{c2+}$ shows that  rattle vibration 
increases but  spin reversal does not occur.
The solid line with the initial spin
$n_{c2+}<n_0=0.005$ shows that the spin reversal behavior
agrees with that discussed in subsection \ref{nc2}. 
Fig. \ref{nc21pm} (b) shows the behavior of $n(t)$ 
for negative initial spins $n_0=-0.006$ (dotted line), $-0.015$ (dashed line),
 and $-0.025$ (solid line).
Even if the initial spin is negative, the same behavior occurs
about $n_{c1-}$ and $n_{c2-}$.
Thus, it is observed that spin reversal behaviors are depended on these
critical values.

\subsection{Number of spin reversals $n_r$ and friction coefficient $\mu$}
This subsection discusses how the  number of spin reversals $n_r$ 
changes depending on the friction coefficient $\mu$.
This number of spin reversals is obtained from a simulation  based on
the exact system described in Eqs. (\ref{Ldot})-(\ref{visf}).
 
\begin{figure}[h]
  \begin{center}
    \includegraphics[width=0.35\textwidth]{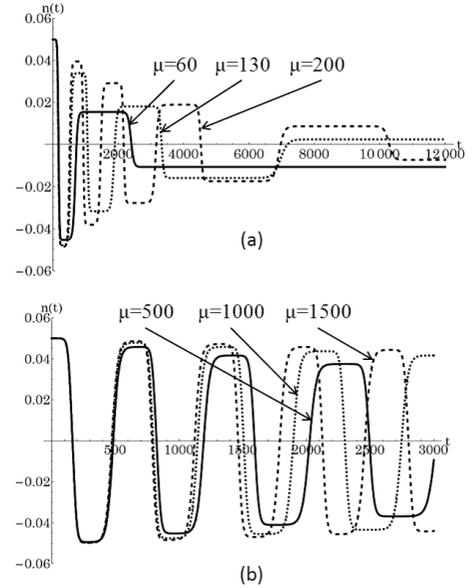}
    \caption{\small{Time evolution of $n(t)$. (a) The initial values of $\mu$ are
    $60$ (solid line), $130$ (dotted line), and $200$ (dashed line).
     (b) The initial values of $\mu$ are
    $500$ (solid line), $1000$ (dotted line), and $1500$ (dashed line).}\label{mu601000}}
  \end{center}
\end{figure}

The initial conditions are $x_0=y_0=0.01$, $\dot{x}_0=\dot{y}_0=0$, and $n_0=0.05$ with 
parameters $a=10$, $b=3$, and $\delta=0.03$.
Fig. \ref{mu601000} (a) shows the behavior of $n(t)$ for
$\mu=60$ (solid line), $130$ (dotted line) and $200$ (dashed line).
It is observed that for each of these $\mu$ values, the number of spin reversals increases to
three, six, and nine times, respectively.
Fig. \ref{mu601000} (b) shows the behavior of spin $n(t)$ for
$\mu=500$ (solid line), $1000$ (dotted line), $1500$ (dashed line). 
Note that the spin behavior is similar to that for the no-slip case 
 as the friction increases.

We usually consider that when friction enlarge, energy loss become large, so that
the reverse number decrease. The numerical simulations show the opposite thing to 
this intuition. 
In reality, the  contact with the horizontal plane is not a point but an area,
spinning torque by rotation $n$ seems to effect the dynamics of rattleback
 as discussed by Garcia and Hubbard\cite{gh}.
The adopted model does not include this spinning torque,
so that energy loss depends only on the velocity $\bvp$ which becomes small when
friction coefficient $\mu$ becomes large, thus, it may be considered that
the number of spin reversal increase.
Moreover, in  reality, when the spin reversal does not occur, the contact point $x_p$ settles
down to $x=y=0$ and $z=1$, and the slip velocity $\bvp$ is equal to zero. 
Then, the rattleback stops spinning after a while.
In contrast, in the numerical simulation, if we set the conditions $\bvp=0$,
$\ddot{x}=\dot{x}=x=0$, and $\ddot{y}=\dot{y}=y=0$, 
$\dot{n}$ becomes zero from Eq. (\ref{I3dotn}).
Therefore, the spin $n(t)$ becomes constant and does not stop after the spin reversal ends,
 as shown in Fig. \ref{mu601000} (a).
It is considered that this phenomena are also depended on not considering spinning torque
due to the frictional force by spinning. 
 
\subsection{The approximate number of spin reversals $n_{r:ap}$ versus 
the exact number of spin reversals $n_{r:ex}$}
In this subsection, the approximate number of spin reversals $n_{r:ap}$ 
obtained from Eqs. (\ref{nr1}) and (\ref{nr2}) in subsection \ref{rn} is compared
with the exact number of spin reversals $n_{r:ex}$ obtained from a simulation  based on
the exact system described in Eqs. (\ref{Ldot})-(\ref{visf}).

\begin{figure}[h]
  \begin{center}
    \includegraphics[width=0.4\textwidth]{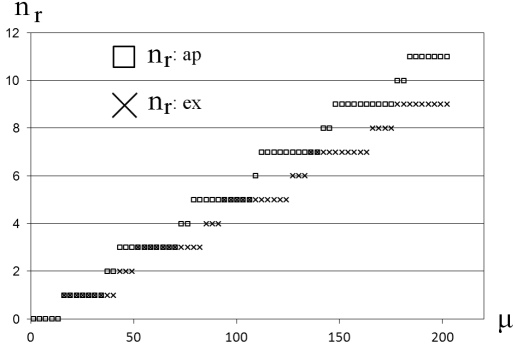}
    \caption{\small{The number of spin reversals $n_r$ versus $\mu$: $\Box$ represents $n_{r:ap}$
    and $\times$ represents $n_{r:ex}$}\label{nrvsmu}}
  \end{center}
\end{figure}

The initial conditions are $x_0=y_0=0.01$, $\dot{x}_0=\dot{y}_0=0$ 
and $n_0=0.05$ with parameters $a=10$, $b=3$, and $\delta=0.03$.
Fig. \ref{nrvsmu} shows the number of spin reversals $n_r$ for a range of 
values of  $\mu$ from $20$ to $200$.
Crosses represent $n_{r:ex}$ and the boxes represent $n_{r:ap}$.
Up to  $\mu \sim 70$, both values are almost identical, but, the difference increases
as $\mu$ increases.
It is assumed that the terms neglected in the approximation affect energy  dissipation.
Thus, for a given value of $\mu$, $n_{r:ap}$ is larger than $n_{r:ex}$.

\subsection{Form factor $L(f)$ and $n_{r:ex}$}
This subsection discusses the relationship between  the form factor 
defined in Eq. (\ref{ff}) and $n_{r:ex}$.

\begin{figure}[h]
  \begin{center}
    \includegraphics[width=0.3\textwidth]{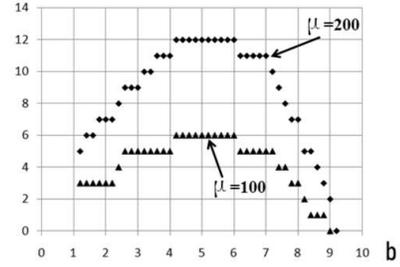}
    \caption{\small{Number of spin reversals $n_{r:ex}$ versus $b$}\label{vsb}}
  \end{center}
\end{figure}

The initial conditions are $x_0=y_0=0.01$, $\dot{x}_0=\dot{y}_0=0$, and $n_0=0.05$ with 
parameters $a=10$, and $\delta=0.03$.
Fig. \ref{vsb} shows the behavior of $n_{r:ex}$ as a function of 
the parameter $b$ for $\mu=100$ and $200$.
Note that the $b$ dependence of $n_{r:ex}$ is similar to that of the form factor
presented in Fig. \ref{lf}. Therefore,
the estimate derived from the effective equations of motion is qualitatively correct. 

%
%
\section{Discussion and conclusions}
%
%
I have examined the behavior of a rattleback with viscous friction to answer
the questions mentioned in the introduction.
The following results were determined analytically and confirmed numerically.

Critical value of the initial spin exist:$n_{c1\pm}$ and $n_{c2\pm}$.
When the initial spin $n_0$ is in the region $n_{c1-}<n_0<n_{c1+}$,
rattle vibration does not increase.
When the initial  spin $n_0$ is in the region $n_{c2-}<n_0<n_{c1-}$
or $n_{c1+}<n_0<n_{c2+}$, rattle vibration increases but spin reversal does not occur.
A numerical simulation based on the exact equations of motion confirmed the existence of
these values. 

I theoretically obtained the number of spin reversals $n_r$ as a function of
the coefficient of friction $\mu$ and the form factor $L(f)$, which contains
the ratio of $a$ to $b$.
From the expression of $n_r$, it was found that the number of spin reversals increases
as $\mu$ increases and a certain value of the ratio of $a$ to $b$ 
gives the maximum number of spin reversals.

In this paper, viscous friction is adopted  to perform a first examination
of spin reversal. However, in reality 
it is assumed that  spin reversal is associated with the effect of Coulomb friction.
Furthermore, it seems that rolling friction and friction by spinning are also effective.
Therefore, these frictions should be adopted to further understand the  rattleback behavior.
This is a subject for future investigations.
\begin{acknowledgments}
I thank members of the elementary particle physics groups at Niigata University and 
Yamagata University for their valuable comments on my study  in an autumn seminar 
at Bandai Fukusima in Japan(2010).
I also appreciate the support received at a workshop hosted 
by the  Yukawa Institute for Theoretical Physics at Kyoto University.
\end{acknowledgments}

\bibliography{rattleback}

\end{document}